\documentstyle[epsfig]{mn2e}

\begin{document}
\title[Role of gas in grand spiral structure]
{Role of gas in supporting grand spiral  structure}
\author[S. Ghosh  and C.J. Jog ]
       {Soumavo Ghosh$^{1}$\thanks{E-mail : soumavo@physics.iisc.ernet.in}, and
        Chanda J. Jog$^{1}$\thanks{E-mail : cjjog@physics.iisc.ernet.in}\\
$^1$   Department of Physics,
Indian Institute of Science, Bangalore 560012, India \\
}

\maketitle

\begin{abstract} 
The density wave theory for the grand-design two-armed spiral pattern in galaxies 
is successful in explaining several observed features.
 However, the long-term persistence of this spiral structure is a serious problem  since the group transport would destroy it within about a billion years as shown 
in a classic paper by Toomre.
In this paper we include the low velocity dispersion component, namely gas, on an equal footing with stars in the formulation of the density wave theory, and obtain the dispersion relation for this coupled system.   
We show that the inclusion of gas makes the group transport slower by a factor of few, thus allowing the pattern to persist longer - for several billion years. Though still less than the Hubble time,
this helps in making the spiral structure more long-lived. 
Further we show that addition of gas is essential to get a stable wave for the observed pattern speed for the Galaxy,
 which otherwise is not possible for a one-component stellar disc.
\end{abstract}


\section{Introduction}
The grand two-armed spiral pattern as seen in galaxies such as M81 or M51 makes a striking visual impression. Although these spiral patterns  have been studied for over five decades, their origin and persistence are still not fully understood.

It was realized early on that a material spiral feature would get wound up in a few rotation periods, due to the differential rotation in a galactic disc. Since spiral features are commonly seen, it was
proposed by several authors starting from B. Lindblad, and others including Lin \& Shu (1964, 1966), that at 
least the grand spiral patterns seen in  spiral galaxies are density waves governed mainly by gravity. In this theory, the  spiral pattern is claimed to be stationary which gets around the winding problem.
Thus the spiral pattern is claimed to last for times much longer than rotational time-scales. Further, Lin, Yuan \& Shu (1969) successfully interpreted some of the observable features of spiral galaxies  by this theory. For a good exposition of the density wave theory, see
Rohlfs (1977).

However, several questions have been raised about the validity of this theory, such as whether or not these density waves
 are truly stationary (Toomre 1969), and whether galaxies
 indeed admit spiral waves as self-consistent modes of oscillation (Lynden-Bell \& Ostriker 1967), see Pasha (2004) for a review of this topic.
Over the years, many additional aspects have been studied that underline the complexity of the spiral structure, see e.g. the reviews by Toomre (1977), Sellwood (2013), and Dobbs \& Baba (2014). Bars have been suggested as a possible mechanisms responsible for the origin of density wave (Sempere \& Rozas  1997; Athanassoula 2012). Numerical simulations (Sanders \& Huntley 1976, Combes \& Gerin 1985) have shown how even a weak barred potential can trigger a spiral perturbation in the gaseous component.
A tidal encounter generally leads to a global $m=2$ spiral pattern (Zaritsky et al. 1993, Struck et al. 2011, Dobbs \& Baba 2014). Many galaxies
also show flocculent spiral arms which can be explained as being  transient, material spiral features that arise due to swing amplification of non-axisymmetric perturbations, as originally proposed  by Goldreich \& Lynden-Bell (1965), also see Toomre (1981).
In many cases both density waves as well as transient, material spiral arms can co-exist which makes the analysis and application to a particular galaxy more complicated. A galaxy may show more than one pattern speed for m=2 (e.g., Gerhard 2011).
Thus, the subject of density wave theory as applied to galaxies is complex.

In this paper, we will focus on  one specific issue, namely the effect of gas on the existence and the group transport of spiral density waves. 
In a classic paper, Toomre (1969) showed that any packet of such waves moves 
radially, and towards  increasingly shorter wavelengths, with a group velocity that is sufficient to destroy the wave packet itself within a few galactic revolutions. This was a major setback to the persistence of "stationary" density waves as proposed by Lin \& Shu (1966). 
Since it is known observationally that regular, grand spiral features are common, 
Toomre argued that the presence of a mechanism to regenerate those grand features is required, such as due to tidal forcing, for the density wave picture to be saved. Meinel \& Ruediger (1987) addressed the question of persistence of the grand spiral structure by means of a wave packet with vanishing group velocity.

A typical spiral also contains a low velocity dispersion component, namely gas, in addition to stars. For a more realistic and complete treatment of galactic dynamics, the dynamical effect of gas needs to be taken into account. The addition of gas is shown to make the disk significantly more unstable as shown for local, axisymmetric perturbations (Jog \& Solomon 1984 a, b; Bertin \& Romeo 1988; Jog 1996 ). In the non-axisymmetric case,  gas  has a strong effect on the resulting swing amplification (Jog 1992), and it is shown to result in broader arms as observed
(Schweizer 1976).

The gas fraction by mass is measured to vary from $\sim$ 4 per cent for Sa type galaxies to $\sim$ 25 per cent for Scd type galaxies (Young \& Scoville, 1991; Binney \& Merrifield 1998).
A study 
using the deep Spitzer survey data (Elmegreen et al. 2011) revealed the trend that early type galaxies tend to have multiple arms and grand design spiral pattern, while the late type galaxies show  mainly flocculent spiral features.
Thus gas fraction may be correlated to the type of spiral structure.
Though the response of gas component to spiral density wave has been studied theoretically (e.g., Roberts 1969) as well as observationally for galaxies like M51    
 (Rand 1993), 
the role of gas in the frame work of density wave theory has so far not drawn the attention it deserves.

In this paper we consider gas along with stars in the formulation of the density wave theory, and obtain the dispersion relation for this coupled system.
Although the response of interstellar gas to the stellar density wave has been well-studied, the role of gas in the group transport has not been considered so far. 
 We find that the addition of gas lowers the group velocity thus allowing the grand design spiral pattern to persist longer.
Further, we show that addition of gas permits waves to be real for the observed pattern speed, which otherwise cannot be realized for a one-component stellar disc.
Thus, the dynamical effect of gas helps spiral arms to exist and persist longer.

Section 2  contains the details of the formulation of the problem, section 3 presents 
the results. Sections 4 and 5 contain the discussion and conclusion respectively.
\section{Formulation of the Problem}
\subsection{Dispersion relation for a two-component disc}
Here we treat a galactic disc as a gravitationally coupled two-component (stars plus gas) system, where the stars are taken to be collisionless  and characterized by a surface density($\Sigma_{0s}$) and a one-dimensional velocity dispersion, $\sigma_s$, and gas as a fluid characterized by surface density $\Sigma_{0g}$ and a one-dimensional velocity dispersion or the sound speed, $\sigma_{g}$.
For simplicity of calculation, the galactic disc is assumed to be infinitesimally thin and pressure acts only in the disc plane i.e. in this paper we are interested in gravitational instabilities in the disc plane only. We use cylindrical co-ordinates $(R, \phi, z$).

We derive the dispersion relation for such a joint system in the WKB limit or the tightly wound case, following the procedure as in Binney \& Tremaine (1987); hereafter BT87.
The small density perturbations are taken to be of type exp $ [ i (\omega t - m \phi + kR) ]$ where $\omega$ is the frequency, and  $k$ is the wavenumber. This simple modal approach assumes a constant pattern speed $\Omega_p = \omega/ m$. However, in a realistic case, the quantities $\omega$ and $k$ could vary gradually with 
radial location and time, as centred around a mode (as in Toomre 1969). 
Here the same dispersion relation as obtained for the modes is taken to be valid to describe the behaviour of these general disturbances that can be used to study wave packets with gradually varying properties (e.g., Whitham 1960, Lighthill 1965).

The dispersion relation is obtained to be
\begin{equation}
\frac{2\pi G \Sigma_{0s} |k|F\Big(\frac{\omega-m \Omega}{\kappa},\frac{k^2\sigma^2_s}{\kappa^2}\Big)}{\kappa^2-(\omega-m\Omega)^2}+\frac{2\pi G \Sigma_{0g} |k|}{\kappa^2-(\omega-m\Omega)^2+\sigma^2_gk^2} = 1
\end{equation}
Here the function $F$ is the reduction factor
with an expression as given in  BT87, also see Appendix A. 
 This factor  physically takes account of the reduction in $\omega^2$ due to the velocity dispersion of stars.

Rafikov (2001) had  obtained the dispersion relation for a system comprised of n distinct collisionless systems along with gas for the axisymmetric case (m = 0, see equation 22 in that paper). Thus the dispersion relation for our case
is given as a special case corresponding to n=1.

Next we define:
\begin{eqnarray}\nonumber
\alpha_s = \kappa^2-2\pi G\Sigma_{0s}|k| F\Big(\frac{\omega-m \Omega}{\kappa},\frac{k^2\sigma^2_s}{\kappa^2}\Big)\\\nonumber
\alpha_g = \kappa^2-2\pi G\Sigma_{0g}|k|+k^2{\sigma}^2_g\\\nonumber
\beta_s = 2\pi G\Sigma_{0s}|k|F\Big(\frac{\omega-m \Omega}{\kappa},\frac{k^2\sigma^2_s}{\kappa^2}\Big)\\\nonumber
\beta_g = 2\pi G\Sigma_{0g}|k| \\
\end{eqnarray}
Then equation (1) reduces to:
\begin{eqnarray}
(\omega-m\Omega)^4-(\alpha_s+\alpha_g)(\omega-m\Omega)^2+(\alpha_s\alpha_g-\beta_s\beta_g)=0
\end{eqnarray}
This is a quadratic equation in $(\omega-m\Omega)^2$. Solving it we get,
\begin{eqnarray}\nonumber
\!\!(\omega-m\Omega)^2=\frac{1}{2}\{(\alpha_s+\alpha_g)\pm [(\alpha_s+\alpha_g)^2-4(\alpha_s\alpha_g-\beta_s\beta_g)]^{1/2}\}\\
\end{eqnarray}
The  additive root for $(\omega-m\Omega)^2$ always lead to a positive quantity, hence it indicates always oscillatory perturbations under all conditions (same as for axisymmetric case, see Jog \& Solomon, 1984a); In order to study the stability of the system and its further consequences, we therefore consider only the negative root which is:
\begin{eqnarray}\nonumber
~~~~~~~~(\omega-m\Omega)^2=\frac{1}{2}\{(\alpha_s+\alpha_g)- [(\alpha_s+\alpha_g)^2-\\\nonumber
~~~~~~~~~4(\alpha_s\alpha_g-\beta_s\beta_g)]^{1/2}\}\\
\end{eqnarray}
Next define two dimensionless quantities, $s$, the frequency, and $x$, the dimensionless wavenumber as:
\begin{eqnarray}
s=({\omega-m\Omega})/{\kappa} =   {m(\Omega_p - \Omega})/{\kappa}, \: \:  x = k / k_{crit}
\end{eqnarray}
\noindent where $k_{crit} = \kappa^2 / 2 \pi G  (\Sigma_{0s} + \Sigma_{0g})$. For one-component case, say with $\Sigma_{0g}=0$, this is the largest stable wavenumber for a pressureless stellar disc.

Dividing both sides of equation (5) by $\kappa^2$ and using the above two dimensionless quantities, we get the dimensionless form of the above dispersion relation as:
\begin{eqnarray}
s^2=\frac{1}{2}[(\alpha'_s+\alpha'_g)-\{(\alpha'_s+\alpha'_g)^2-4(\alpha'_s\alpha'_g-\beta'_s\beta'_g)\}^{1/2}]
\end{eqnarray}
where,
\begin{eqnarray}\nonumber
\alpha'_s=1-(1-\epsilon)|x|F(s,\chi) \\\nonumber 
\alpha'_g=1-\epsilon |x|+\frac{1}{4}Q^2_g\epsilon^2x^2 \\\nonumber
\beta'_s=(1-\epsilon)|x|F(s,\chi) \\\nonumber
\beta'_g = \epsilon |x|\\
\end{eqnarray}
where, $\chi$= ${k^2\sigma^2_s}/{\kappa^2}$ = $0.286 Q_s^2 (1-\epsilon)^2 x^2$.  
The three dimensionless parameters $Q_S$, $Q_g$ and $\epsilon$ are respectively the Toomre $Q$ factors for stars as a collisionless system  $Q_s$(=$\kappa \sigma_s /(3.36 G \Sigma_{0s})$), and for gas $Q_g$ = ($\kappa \sigma_g /(\pi G \Sigma_{0g})$)
and;  $\epsilon$ =${\Sigma_{0g}}/( {\Sigma_{0s}+\Sigma_{0g}})$ the gas mass fraction in the disc respectively.
Similarly, the one-component analog of this dispersion relation is (BT 87):
\begin{equation}
s^2= 1-|x|F\Big(\frac{\omega-m \Omega}{\kappa},\frac{k^2\sigma^2_s}{\kappa^2}\Big)
\end{equation}

\subsection {Group velocity for a two-component disc}
It is well-known that information from a disturbance, generated at a given radius, propagates radially with its group velocity (e.g., Whitham 1999).
For an inhomogeneous medium, the group velocity at a fixed radius is defined as (Whitham 1960, Lighthill 1965):
\begin{equation}
c_g(R)= \frac{\partial \omega(k,R)}{\partial k}
\end{equation}

Here we study effect of gas on the radial group velocity
of a wave packet in a gravitationally coupled star-gas system. 
From the dispersion relation for a collisionless case, it can be seen that evaluating an analytic expression for the group velocity is cumbersome due to the implicit form of the reduction factor. This is particularly so for the two-component case.
However, interestingly, the group velocity can be easily estimated graphically from the slope of $s$ versus. $x$ (see Fig.1, Sec. 3), and
 is given as (Toomre 1969; BT 87):
\begin{equation}
c_g(R)= sgn(ks)(\kappa/k_{crit}) \frac{ds}{dx}
\end{equation}
\noindent where $sgn(ks) = \pm 1$ depending on whether $ks >0$ or is $< 0$.
This approach was developed by Toomre (1969), we apply it here for the frequency $s$ obtained above for the two-component case (eq. [7]).
\section{Results} 
We next investigate how the addition of gas affects the group velocity and hence the radial group transport of a wave packet. 
Driven by purely theoretical interest, here we carried out the analysis at a fixed radius in the disc varying the gas surface density from 5 to 20 per cent of the stellar surface density. For a real galaxy, at a particular radius one would expect to get a unique ratio of gas to stellar surface density, nevertheless our approach would 
throw some light on what is likely to happen at those radii for which the observed gas fraction matches with the values considered here in the analysis.
We will compare our results with those of Toomre (1969), the latter obtained for the stars-alone case, with $Q_s$=1. For a two-component case, when $Q_s=1$, the system will
 become unstable (Jog \& Solomon 1984a), and the group velocity concept is not applicable in that case.
Hence we consider a slightly higher $Q_s$ = 1.3 since that leaves the two-component system stable even for higher gas fraction of $20\%$ (Jog 1996).
For given $Q_s$ and $\epsilon$ values, and assuming a certain value for the ratio of dispersions $\sigma_s$ to $\sigma_g$ (taken to be = 3.5 as observed in the solar neighbourhood, see Narayan \& Jog 2002), one can obtain $Q_g = (0.306 Q_s)(1-\epsilon)/\epsilon$. Using these as input parameters, the values of $s$ versus. x can be obtained for different gas fraction values.

In Fig. 1, we plot the dispersion relation in its dimensionless form 
for different gas fractions (eq. [7]), including for the stars-alone case (eg. [9]).
\begin{figure}
\centering
\includegraphics[height=2.2in,width=3.0in]{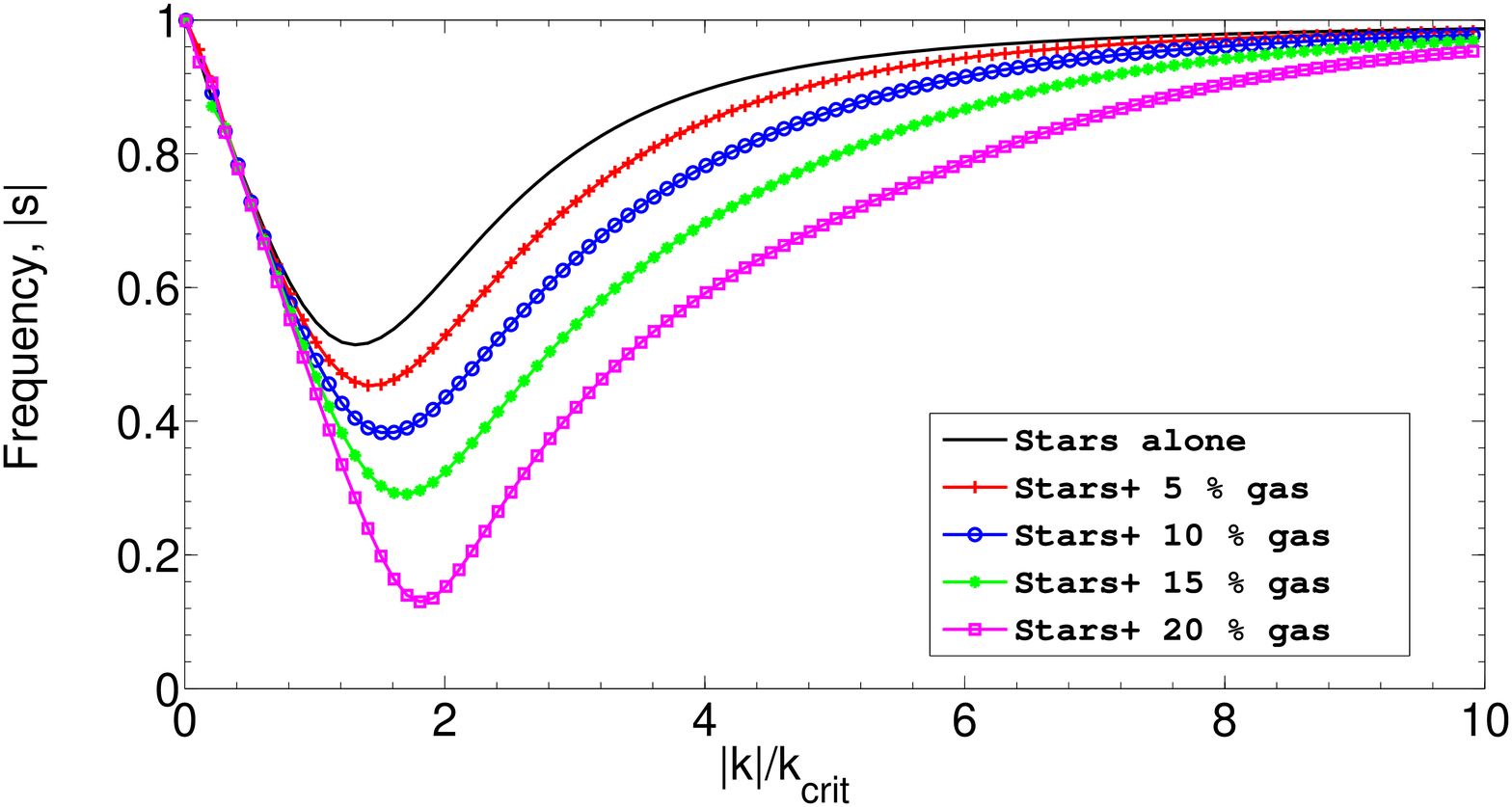}
\caption{Dispersion relation(eq. 7 \& 9), plotted in its dimensionless form. 
Since the dispersion relation is symmetric with respect to both s and x, only their absolute
values are shown in this figure. As the fraction of gas is increased, the system gets closer to being unstable.}
\end{figure}
\noindent Two points are clear from this figure.
First, as more gas is added, the plot for dispersion relation progressively occupies more of the forbidden region between corotation ($s$=0) and inner Lindblad resonance(ILR) ($s$ =-1), implying that the system becomes more and more prone to being unstable, along the same line as seen for the axisymmetric case (Jog \& Solomon 1984a). 
Secondly, a careful inspection also reveals that in the relatively short branch (high wavenumbers) of the dispersion relation, the plots become moderately flatter as more gas is added to the system.

While the first feature is somewhat expected and relatively well understood, the second point is particularly interesting. As the group velocity is directly related to the slope of these curves (see Section 2), these curves give a clear hint  that when gas is included, the radial group velocity of a wave packet whose central wavenumber falls in the region, is likely to decrease, though moderately.

To investigate this issue further, we estimated the value of group velocity at R= 8 kpc for different gas fractions.
The values of $\kappa$ and $\Omega$ are calculated from a flat rotation curve with a constant radial velocity of 220 km s$^{-1}$ at $R=8 kpc$ near the solar neighbourhood in the Galaxy, obtained from the standard mass model by Mera et al. (1998). 
These are: $\kappa$ = 38.9 km s$^{-1}$ kpc$^{-1} $, and $\Omega$ = 27.5 km s$^{-1}$ kpc$^{-1}$. The pattern speed $\Omega_p$ for $m=2$ is taken to be 12.5 km s$^{-1}$ kpc$^{-1}$ as in Toomre (1969), so as to ensure that any variation in results is due to the inclusion of gas. The value of the slope is obtained graphically from the curve of s versus. x (Fig. 1) and is calculated at a point x where the line $s$ = constant (for a particular pattern speed) intersects the curve. The value of $x$ corresponding to the higher $x$ or $k$ value is chosen rather than the lower $x$ value since the tight-winding approximation is better satisfied at the shorter wave branch. This gives the value of the group velocity (eq.[11]), see Table 1 for the results. 
\begin{table}
\centering
  \begin{minipage}{140mm}
   \caption{Group velocity for various gas fraction values}
\begin{tabular}{lllll}
\hline
$\Sigma_{0g}/\Sigma_{0s}$   & Slope & Group velocity  & Time to travel \\%
&  &  (km /sec)& 10 kpc  (Gyr) \\
\hline
0.0 &  0.14 & 8.0 & 1.2\\
0.05 &  0.11 & 6.0  & 1.6\\
0.10 &  0.10 & 5.5 & 1.7\\
0.15 &  0.08 & 4.4 & 2.2 \\
0.20 &  0.07 & 4.0 & 2.4 \\
\hline
\end{tabular}
\end{minipage}
\end{table} 
As the gas fraction increases from 0 to 20 per cent,  the group velocity is reduced by a factor of two. Consequently, the time taken by a wave packet to travel a distance of 10 kpc is about two times longer. Even these rough estimates tell us that 
the addition of gas helps the density pattern to persist for a relatively longer time-scale. This is the main finding of this work.
We stress that the quantitative result is not robust as it depends on the slope of the curve measured graphically. For example, for a slightly higher pattern speed of 14 km s$^{-1}$ kpc$^{-1}$, the group velocity has a range of 11 to 4.4 km s$^{-1}$ as the gas fraction is increased from 0 to 20 per cent, hence the time taken for a wavepacket to travel the same distance of 10 kpc increases almost by a factor of three. Thus the results in Table 1 give a typical
sense of increase in the lifetime of a spiral pattern on including gas in the picture.

The modern observations of the pattern speed for the Milky Way show higher values, that lie in a range between 17-28 km s$^{-1}$ kpc$^{-1}$ (e.g. Gerhard 2011, Siebert et al. 2012, Junqueira et al. 2015), with a typical value of $\Omega_p$ = 18 km s$^{-1}$ kpc$^{-1}$. For most of this range of values for $\Omega_p$, the value of $s$ in the middle range of the Galaxy is $< 0.5$. For this value of $s$, the dispersion relation (eq.[9]) does not admit a real solution for $Q_s = 1.3$.
Instead its solution  has an imaginary wavenumber, and hence it decays exponentially with radius on scales of $\sim$ a few $\lambda_{crit}$(BT 87, Chapter 6.2). 
For the solar neighbourhood, with $(\Sigma_{0s}+\Sigma_{0g})$ = 52 M$_{\odot}$ pc$^{-2}$ (Narayan \& Jog 2002), the value of $\lambda_{crit}$ is $\sim$  6 kpc. 
This aspect does not seem to be recognized or at least mentioned in the papers which give the observational determination of the pattern speed.
Thus, a one-component stellar disc is not adequate to describe the observed pattern, and therefore taking account of gas is essential in order for the  density waves to be stable as shown next.

For the recent observed value of the pattern speed of 18 km s$^{-1}$ kpc$^{-1}$, the value of the frequency is obtained as $s=0.5$. For this value of $s$,
the wavenumber for the $s$ versus. x curve for  the stars-alone case would be imaginary (see fig. 1). Note, however, that in contrast,
for gas fractions of 10 to 15 per cent, the curve $s$ versus. x does have real solutions for the wavenumber and the slope at these wavenumbers 
remains  nearly constant $\sim 0.2 $. The corresponding group velocity is  11 km s$^{-1}$,
comparable to but higher than the values in Table 1. Thus the time to travel 10 kpc is $\sim 10^ 9$ yrs.
Thus,  interestingly, we find that for the observed pattern speed in the Galaxy, the waves for stars-alone case have an imaginary wavenumber, hence are evanescent. For a stable wave solution one needs to take account of gas.  
Moreover, addition of gas allows somewhat higher pattern speeds which are consistent with observations, to be valid in a galaxy. Since the pattern speed decides the location of resonance points, the inclusion of gas thus allows the corotation to be shifted to an inner radius.\\
\section{Discussion}
\subsection{Stars: collosionless versus. fluid approach}
In this paper we have treated stars as a collisionless system involving no pressure term, as required for a correct treatment of the group velocity.
There are some problems in dynamics where the result does not depend critically on whether a fluid or a collisionless representation is used for stars. For example, 
in the local stability analysis of axisymmetric perturbations in a galactic disc leading to the Toomre Q criterion, results obtained from both approaches 
match quite satisfactorily. The Toomre
$Q$ factors (Toomre 1964) for stars and gas differ by only 7 per cent, with a factor 3.36 replacing $\pi$ in the denominator for a collisionless case (e.g., BT 87).

However, it is known that the two approaches differ substantially with the system being more bouncy at high wavenumbers in the fluid case as can be seen by the plot of the dispersion relation in the two cases (e.g., Fig. 6.14, BT 87; also see Rafikov 2001). We plot the dispersion relation with 15 per cent  gas for two cases: first stars as a collisionless system (eq. [8] above; and then treating stars as a fluid (with the dispersion relation as in Jog \& Solomon 1984a), see Fig. 2.
\begin{figure}
\centering
\includegraphics[height=2.4in,width=3.2in]{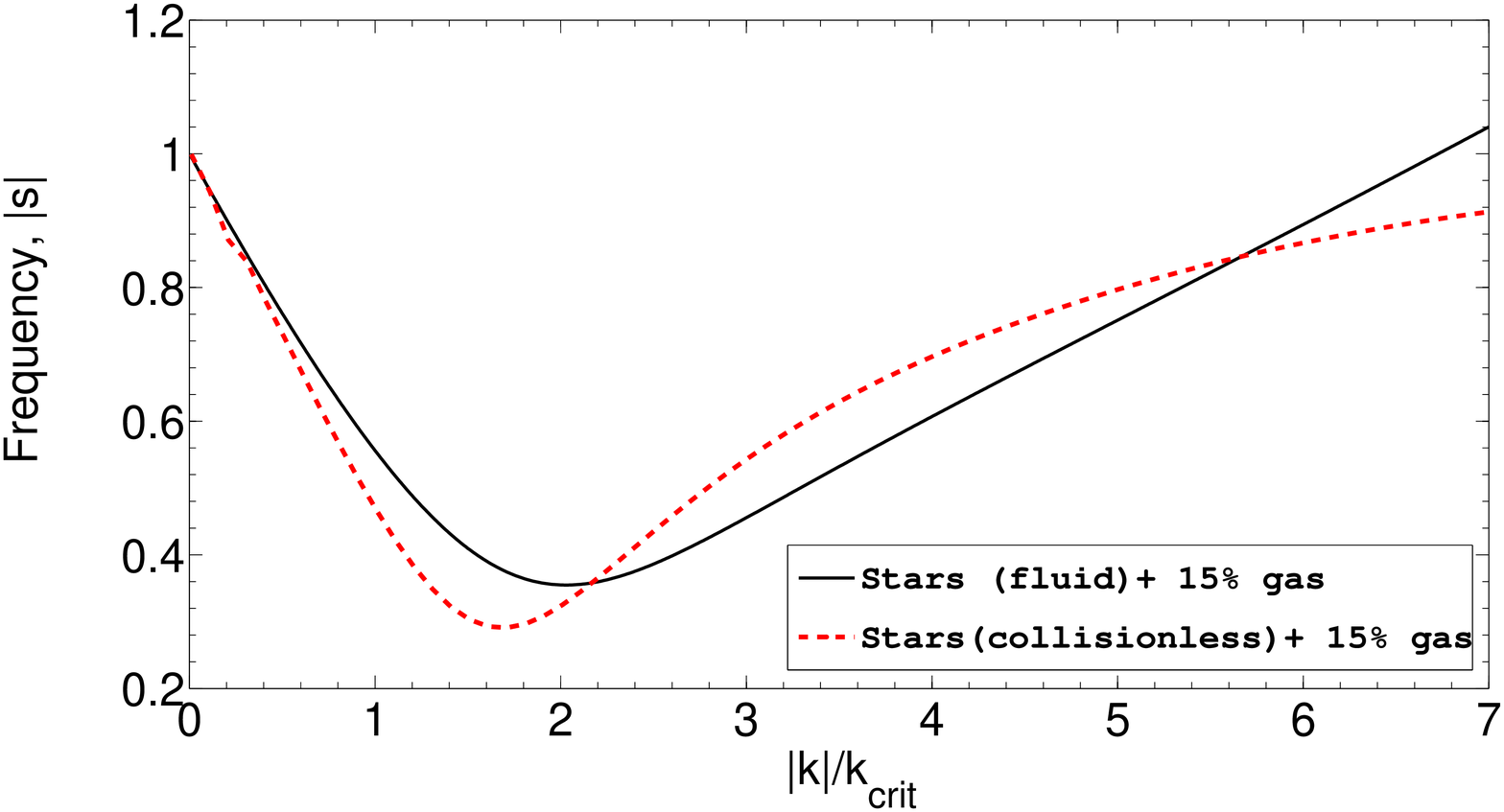}
\caption{Dispersion relation(eq. 7 \& 9), plotted in its dimensionless form,  
at R = 8 kpc for a disc with 15 per cent gas,  for
two cases: first treating stars as a collisionless system and then treating stars as a  fluid. The two approaches agree at low $x$ values, but differ at large $x$ values.}
\end{figure}
From Fig. 2, it is  evident that for the case where the stars are treated as a fluid, the curve for the dispersion relation is steeper than the case where stars are treated as collisionless system and the $s$ values exceed 1 beyond x=7. In contrast, in the collisonless treatment, the $s$ values are always less than 1 and saturate to 1 at x $>> 7$. Since the technique adopted here to calculate group velocity is strongly dependent on slope of these curves, consequently these two different approaches, using the same gas fraction, are bound to give different estimates for the group velocity.
\subsection{Other Issues:}

We would like to caution that, from our result it may appear that with the inclusion of gas by more than 20 per cent or so which is the extreme case considered for this work, may still improve the persistence of the spiral features. But in a real gas-rich galaxy, other processes  like swing amplification generating local flocculent spiral features do take place simultaneously. In particular, for gas rich galaxies it is probably the swing amplification which supersede the grand spiral structure, as evident from the abundance of more flocculent spiral features found in gas rich Sc type galaxies. 

Another point to note is that the analysis presented here is based on a linear perturbation theory so that the non-linear effects have been neglected. But these non-linear effects could have far-reaching consequences. 
A recent study using high- resolution N-body simulations (D'Onghia, Mark \& Hernquist 2013) showed that the nonlinear effects can significantly modify the process of origin and persistence of flocculent spiral features, produced through swing amplification. So, it is worth checking the significance of nonlinearity in the frame work of density wave theory with gas treated on equal footing with stars.
\section{conclusion:}
In summary, we have studied the effect of gas by treating a galactic disc as a two-component gravitationally coupled stars plus gas
system in the framework of density wave theory. The resulting frequency versus. wavenumber curve is flatter which leads to a
lower group velocity.  
The idea of group velocity as an indicator of transport of information is 
routinely studied in other branches of physics, such as quantum mechanics, however its usefulness is underutilized in astrophysics. In the case of 
galactic discs, this idea was applied by Toomre (1969) to show that the density waves cannot last for more than $\sim$ 10$^9$ yrs. 
We have shown that taking account of gas lowers the group velocity by a factor of few, which allows the 
density waves to last longer, to about few $10^9$ yr. This helps persistence of the spiral structure for a longer time-scale in a gas-rich galaxy,
irrespective of the mechanism, such as tidal interaction, that  gives rise to the grand two-armed spiral structure.

The second important result from this work is that for the observed pattern speed of $\sim 18$ km s$^{-1}$ kpc$^{-1}$ for the Galaxy (Siebert et al. 2012), 
the solution gives an evanescent wave (i.e., a wave with an imaginary wavenumber) for a one-component stellar disc, hence the wave would disperse radially. 
We show that it is the inclusion of gas that makes it possible to have a stable wave for the measured pattern speed.  
The addition of gas thus allows somewhat higher pattern speeds to be valid in a galaxy
which otherwise cannot be realized for a one-component stellar disc. Since the pattern speed decides the location of resonance points, the addition of gas thus allows the corotation to be shifted to an inner radius.
This could have implications for the angular transport properties due to the spiral pattern.

\section{Acknowledgements}
C. J. J thanks DST, Govt. of India for support through a J. C. Bose Fellowship.

\bigskip

\noindent {\bf References}

\medskip

\noindent Athanassoula, E. 2012, MNRAS, 426, L46

\noindent Bertin, G., \& Romeo, A. 1988, A\&A, 195, 105

\noindent Binney,J., \& Merrifield, M. 1998, Galactic Astronomy. Princeton Univ. Press, Princeton, NJ 

\noindent Binney, J., Tremaine, S., 1987, Galactic Dynamics. Princeton
Univ. Press, Princeton, NJ (BT87)

\noindent Combes, F., \& Gerin, M., 1985, A\&A, 150, 327

\noindent Dobbs, C., \& Baba, J. 2014, PASA, 31, 35

\noindent D'Onghia, E., Mark, V., \& Hernquist, L., 2103, ApJ, 766, 34

\noindent Elmegreen, D. M. et al., 2011, ApJ, 737, 32

\noindent Gerhard, O. 2011,Mem. Soc. Astron. Ital. Suppl., 18, 185

\noindent Goldreich, P., \& Lynden-Bell, D. 1965, MNRAS,130, 125

\noindent Jog, C.J., \& Solomon, P.M. 1984 a, ApJ, 276, 114

\noindent Jog, C.J., \& Solomon, P.M. 1984 b, ApJ, 276, 127

\noindent Jog, C.J. 1992, ApJ, 390, 378

\noindent Jog C.J. 1996, MNRAS, 278, 209

\noindent Junqueira, T.C., Chiappini, C., Lepine, J.R.D., Minchev, I., \& Santiago, B.X., 2015, MNRAS,449, 2336 

\noindent Lighthill, M.J. 1965, J. Inst. Math. Appl., 1, 1

\noindent Lin, C. C., \& Shu, F. H., 1964, ApJ, 140, 646

\noindent Lin, C. C., \& Shu, F. H., 1966, Proc. Natl. Acad. Sci. USA, 55, 229

\noindent Lin, C.C., Yuan, C., \& Shu, F. H., 1969, ApJ, 155, 721

\noindent Lynden-Bell, D., \& Ostriker, J.P. 1967, MNRAS, 136, 293

\noindent Meinel, R., \& Ruediger, G., 1987, Ap\&SS, 138, 147

\noindent Mera, D., Chabrier, G., \& Schaeffer, R., 1998, A\&A, 330, 953

\noindent Narayan, C. A., \& Jog, C. J., 2002, A\&A, 394, 89

\noindent Pasha, I.I. 2004, preprint: astro-ph/0406143

\noindent Rafikov, R. R., 2001, MNRAS, 323, 445

\noindent Rand, R. J., 1993, ApJ, 410, 68 

\noindent Roberts, W. W., 1969, ApJ, 158, 123

\noindent Rohlfs, A. 1977, Lecture Notes on Density Wave Theory, Springer-Verlag, Berlin

\noindent Sanders, R. H., \& Huntley, J. M., 1976, ApJ, 209, 53

\noindent Schweizer, F. 1976, ApJS, 31, 313

\noindent Sellwood, J. 2013, in Planets, Stars and Stellar Systems Vol. 5, Eds. T.D. Oswalt and G. Gilmore, , Springer Science+Business Media, Dordrecht, p. 923

\noindent Sempere, M. J., \& Rozas, M., 1997, A\&A, 317, 405

\noindent Siebert, A. et al. 2012, MNRAS, 425, 2335  

\noindent Struck, C., Dobbs, C. L., \& Hwang, J.-S. 2011, MNRAS, 414, 2498

\noindent Toomre, A. 1964, ApJ, 139, 1217

\noindent Toomre, A., 1969, ApJ, 158, 899

\noindent Toomre, A. 1977, ARA\&A, 15, 437

\noindent Toomre A., 1981, in Lynden-Bell D., Fall S. M., eds, The structure and evolution of galaxies. Cambridge Univ. Press, Cambridge, p. 111

\noindent Tremaine, S.D. 2001, AJ, 121, 1776

\noindent Young, J. S ., \& Scoville, N. Z., 1991, ARAA, 581, 625

\noindent Whitham, B. , 1960, J. Fluid Mech., 9 , 347

\noindent Whitham, B. 1999, Linear and Non-linear Waves. Wiley, New York

\noindent Young J. S., Scoville N. Z., 1991, ARA\&A, 581, 625

\noindent Zaritsky, D., Rix, H.-W., \& Rieke, M., 1993, Nature, 274, 123
\appendix
\section{Calculation of the Reduction Factor}
The dispersion relation for collisionless stellar disc in the tight-winding limit is (see eq.[9]):
\begin{equation}
s^2= 1-|x|F(s, \chi)
\end{equation}
\noindent where the reduction factor $F(s, \chi)$ is given by (BT87):\\
\begin{equation}
F(s, \chi)=\frac{2}{\chi}exp(-\chi)(1-s^2)\sum_{n=1}^\infty\frac{I_n(\chi)}{1-\frac{s^2}{n^2}}
\end{equation}
\noindent where $I_n (\chi)$ is a modified Bessel function of order n.
Now the explicit dependence of reduction factor  $F(s, \chi)$ on $s$ makes the dispersion relation (A1) an implicit relation which then has to be solved in a self - consistent manner. For obtaining the solution numerically, one has to truncate the infinite sum after a finite terms in such a way so that the solution is not affected by the truncation process. In other words, if the solution with k+1 terms of the series matches well (within the pre-defined tolerance limit) with the solution having k terms of the series, one can safely truncate the series after k terms.

In some special cases, for example while studying the $m=1$ slow modes in Keplerian discs, Tremaine(2001) showed that 
the reduction factor in the dispersion relation simplifies to one containing only the first term of the infinite sum, hence becomes an explicit relation in s and x(= $k/k_{crit}$)(see eqn. [12] there). But we caution that, in general, such a simplification is not always valid and one has to incorporate more terms of the series to get the actual solution.

Here we illustrate how incorporating further terms of the infinite sum will affect the dispersion relation for a collisionless stellar disc for Q =1.3.
\nopagebreak
\begin{figure}
\centering
\includegraphics[height=2.5in,width=3.5in]{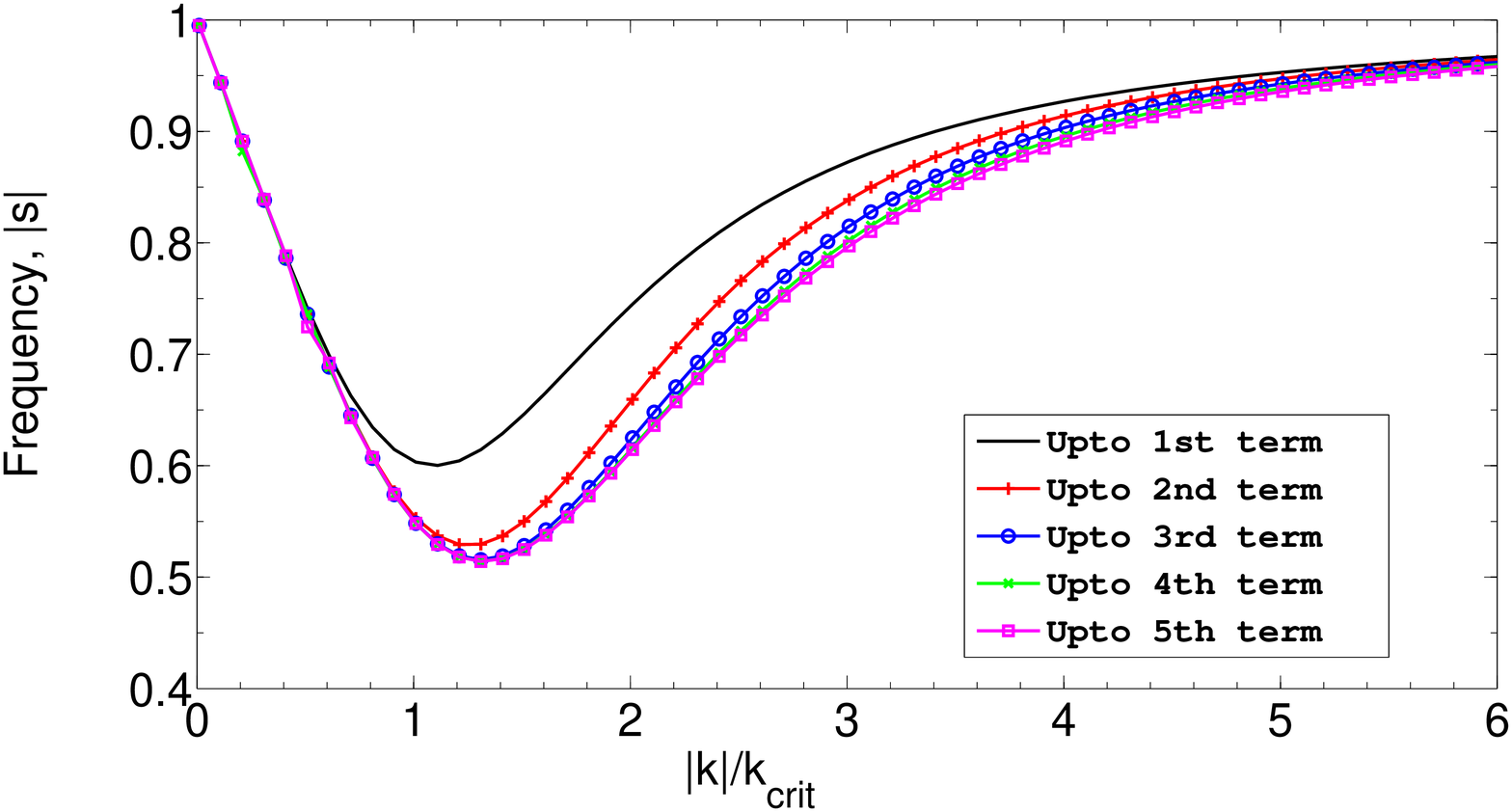}
\caption{Variation of dispersion relation for collisionless stellar disc with Q =1.3 while incorporating subsequent terms of the infinite series in the expression for the reduction factor (eq. [A1]). 
The solution with first four terms matches well with that having first five terms to within 1 per cent, so the series can safely be truncated after the first four terms.}
\end{figure}
From Fig. A1, it is  seen  that with the inclusion of subsequent terms, the solution changes significantly. Allowing more terms of that infinite sum into the solution makes the forbidden region between corotation(s =0) and Inner Lindblad resonance(ILR, s=-1) reduce up to a certain limit, after that the effect gets saturated. This happens for
$n=4$. Hence in our calculation (Section 2), we include terms up to $n=4$ in the reduction factor $F$ as suggested by this figure. The result shown in Fig. A1 agrees with that in Toomre (1969).

\end{document}